\documentclass[aps,prd,twocolumn,nofootinbib,superscriptaddress]{revtex4-2}
\usepackage{graphicx}% Include figure files
\usepackage{dcolumn}% Align table columns on decimal point
\usepackage{bm}% bold math
\usepackage{epstopdf}
\usepackage[backref]{hyperref} 
\hypersetup{colorlinks,%				% Reference color setup	
	linkcolor=blue,%
	citecolor=blue,
        urlcolor =blue}
\usepackage{ulem}
\usepackage{amsmath}
\usepackage{color}
\begin{document}

\preprint{APS/123-QED}

\title{Merging L-shaped resonator with Michelson configuration for kilohertz gravitational-wave detection}% Force line breaks with \\

\author{Xinyao Guo}
\affiliation{Department of Physics, Tsinghua University, Beijing 100084, China}
\email{guoxy24@mails.tsinghua.edu.cn}
\author{Teng Zhang}
\affiliation{School of Physics and Astronomy, and Institute for Gravitational Wave Astronomy,University of Birmingham, Edgbaston, Birmingham B15 2TT, UK}
%\email{tzhang@star.sr.bham.ac.uk}
\author{Denis Martynov}
\affiliation{School of Physics and Astronomy, and Institute for Gravitational Wave Astronomy,University of Birmingham, Edgbaston, Birmingham B15 2TT, UK}
%\email{d.martynov@bham.ac.uk}
\author{Haixing Miao}
\affiliation{Department of Physics, Tsinghua University, Beijing 100084, China}
\affiliation{Frontier Science Center for Quantum Information, Beijing 100084, China}
 \email{haixing@tsinghua.edu.cn}

\begin{abstract}
Detection of gravitational waves in kilohertz frequency range is crucial for understanding the physical processes of binary neutron star mergers. In Ref. [Phys. Rev. X {\bf 13}, 021019 (2023)], a new interferometric configuration has been proposed, employing an L-shaped optical resonant cavity as arm cavity. This alteration enhances the detector's response to kHz signals. However, the departure from conventional Michelson configuration necessitates a redesign of its sensing and control scheme, which is currently under study. In this article, we propose replacing linear arm cavities in the conventional Michelson by the L-shaped resonator. This hybrid configuration features an enhanced  
response at kHz while retaining the same sensing and control scheme as the Michelson setup. At the conceptual level, it exhibits higher sensitivity in the 2-4 kHz range compared to existing configurations.

\end{abstract}

\maketitle

%\tableofcontents

\section{Introduction}\par

\quad Ground-based interferometric gravitational wave (GW) detectors have experienced rapid development in recent years, successfully captured numerous compact binary events\,\cite{Rana, Aasi2015c, Acernese_2015, Kagra, GW170817, GWTC}. Inspired by the binary neutron star merger event GW170817, the scientific community has aimed to enhance detector sensitivity in the kilohertz range to better detect the post-merger signals. However, the high-frequency performance of the traditional Michelson configuration with linear arm cavities is actually limited. At higher optical resonances of arm cavity, the antenna response is partially canceled due to time-averaged effects, making it not the optimal for high frequency detection.

To overcome this issue, many proposals have explored modifications to the traditional Michelson\,\cite{Miao2018, Martynov2019, Ackley2020, Mikhail2019, Page2021, Zhang2021, Wang2022}. The most prominent approach is tuning the system into a signal recycling regime, which involves increasing the finesse of the signal recycling cavity and shifting the peak optical gain to the resonant frequency of coupled SRC-arm cavity\,\cite{TunedAdv}. This idea can effectively improve the detection sensitivity for high-frequency signals, however, the peak sensitivity is typically outside the arm cavity bandwidth, and kilohertz signals remain largely unamplified in the arm, which means any optical loss after the arm cavity strongly limits the high-frequency sensitivity.\cite{Miao2019, Korobko2023}.

Recently, a new configuration with an L-shaped optical resonator as the arm cavity has been proposed \cite{new.configuration}, which shares the same core resonator as the Fox-Smith interferometer \cite{Fox_Smith, PhysRevD.Fox}. By folding the arm cavity into an L-shape, this design directly amplifies gravitational wave signals at the first free spectral range, thereby achieving good high-frequency performance with relatively low SEC losses. However, as the central part is no longer a Michelson interferometer, the sensing and control scheme needs to be redesigned; one such study has been reported in Ref.\,\cite{GuoXY_2023}. 

\begin{figure}[t!]
\centering
\includegraphics[scale=0.41]{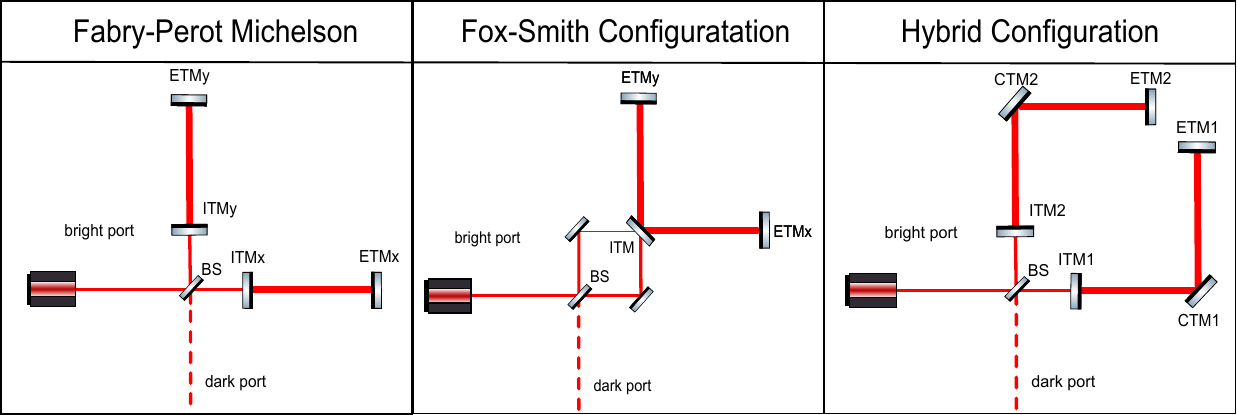}
\caption{A simplified optical layout of the core part of Michelson configuration (left), Fox Smith configuration (middle) and the hybrid configuration (right). The power- and signal-recycling parts are omitted for highlighting. BS: beam splitter, ITM: input test mass, ETM: end test mass, CTM: corner test mass.}
\label{fig:config}
\end{figure}

We realise that the high frequency response of L-shaped resonator is an intrinsic properly of its resonance at the free spectral range, and does not depend on how it is optically coupled. Based on this observation, we propose a hybrid configuration by replacing linear cavities in traditional Michelson configuration with the L-shaped resonators, as illustrated in Fig.\,\ref{fig:config}. 
Not only it can achieve an enhanced kHz response, but also can leverage the sensing and control scheme for Michelson.

The paper is organized as follows: in Sec.\,\ref{sec:overview}, we discuss the principle underlying the hybrid configuration. In Sec.\,\ref{sec:opt}, we discuss the fundamental quantum and classical noise of hybrid configuration, and provided a conceptual noise budget of the hybrid configuration with parameters similar to Ref.\,\cite{new.configuration}. Finally, we conclude our study and provide an outlook in Sec.\,\ref{sec:conclusion}.

\section{Principle}\label{sec:overview}
% optical response, antenna response, total response
%

\quad In this section, a brief illustration of the working principle of hybrid configuration is presented. To illustrate the physics in the Hybrid design more clearly, we selected the special case of normal incidence and plus polarization mode for simplicity [refer to \,\ref{sec:appendix B} for the general case].

Our work basically relies on a conventional decomposition of interferometric gravitational wave detector. A configuration’s detecting sensitivity to GW signals can be decoupled as 2 parts: the antenna response, and the optical response.
The design of the hybrid configuration is actually inspired by an intuitive idea of merging the optical response of the conventional Michelson configuration with the high antenna response of the L-shaped resonator for kilohertz gravitational waves, and thus partially retains the characteristics of both configurations. For clarity, we will illustrate the distinctions and commonalities between hybrid configuration and the Michelson and Fox-Smith configuration from the perspective of antenna response and optical response.

We first discuss the difference in the antenna response between three configurations, which mainly comes from the the different shape of arm cavity. 
The antenna response of configuration is basically the efficiency of gravitational wave signal converting to displacement of the shortest detectable optical length.This factor could be derived by summing up the additional length of optical path introduced by the gravitational wave , as light travels from the point of incidence to the nearest output. The strain to displacement transfer function for traditional Michelson with linear arm cavities of length $L$ could be then analytically expressed as:
\begin{equation}
{\cal B}_{\rm mich}(f) =  L \,{\rm sinc}\left(\frac{2\pi f L}{c}\right) \equiv L\, {F}_{\rm mich}(f)\,,
\end{equation}
where $f$ is the frequency of gravitational wave signal,  $L$
is the total length of arm cavity, and $F_{\rm mich}=F_{+,\rm mich}$ is the normalised antenna responsefor $h_+$ mode in traditional Michelson configuration. As frequency of gravitational wave signal approaches the cavity's free spectral range, the antenna response of a linear cavity to gravitational waves rapidly decreases, limiting the detecting sensitivity of traditional Michelson configuration at high frequencies.

\begin{figure}[t!]
\centering
\includegraphics[scale=0.65]{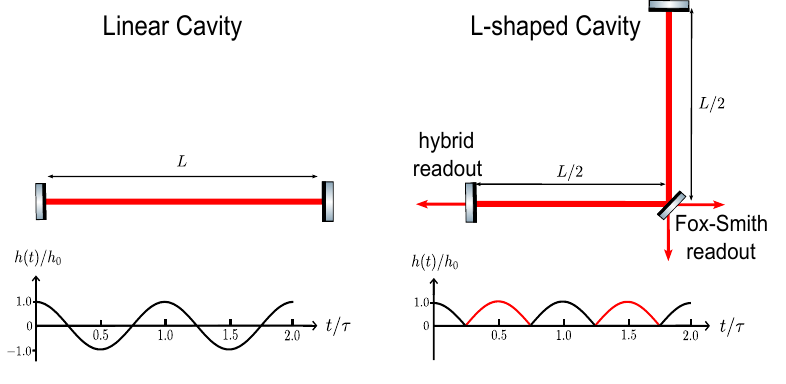}
\caption{Time-dependent response to $h_{+}$-polarized gravitational wave signal with frequency $f=f_{FSR}$ of linear cavity (left) and folded cavity (right), where time is normalized by the round-trip propagation time $\tau=2L/c$. The averaged response over a round-trip propagation period is defined as the antenna response of the configuration. Near the first free spectral range, the periodic sign flipping of the L-shaped cavity results in a heightened antenna response to gravitational wave signals. Meanwhile, for signal away from the first free spectral range, the antenna response of L-shaped cavity also depends on how it is coupled to the output, or equivalently, where the readout is obtained.
}
\label{fig:antenna}
\end{figure}

Folding the arm cavities is one approach to this issue\cite{new.configuration}. For plus mode gravitational signal, L-shaped folding of the arm cavity intricately corresponds to a sign inversion of gravitational wave signal each half round-trip propagation period, making the antenna response for gravitational wave signal at the first free spectral range no longer null response. Fig.\,\ref{fig:antenna} illustrates the effect of periodic sign flipping induced by cavity folding. 

\begin{figure}[htbp]
\centering

    \begin{minipage}[t]{0.95\linewidth}
        \centering
        \includegraphics[width=\textwidth]{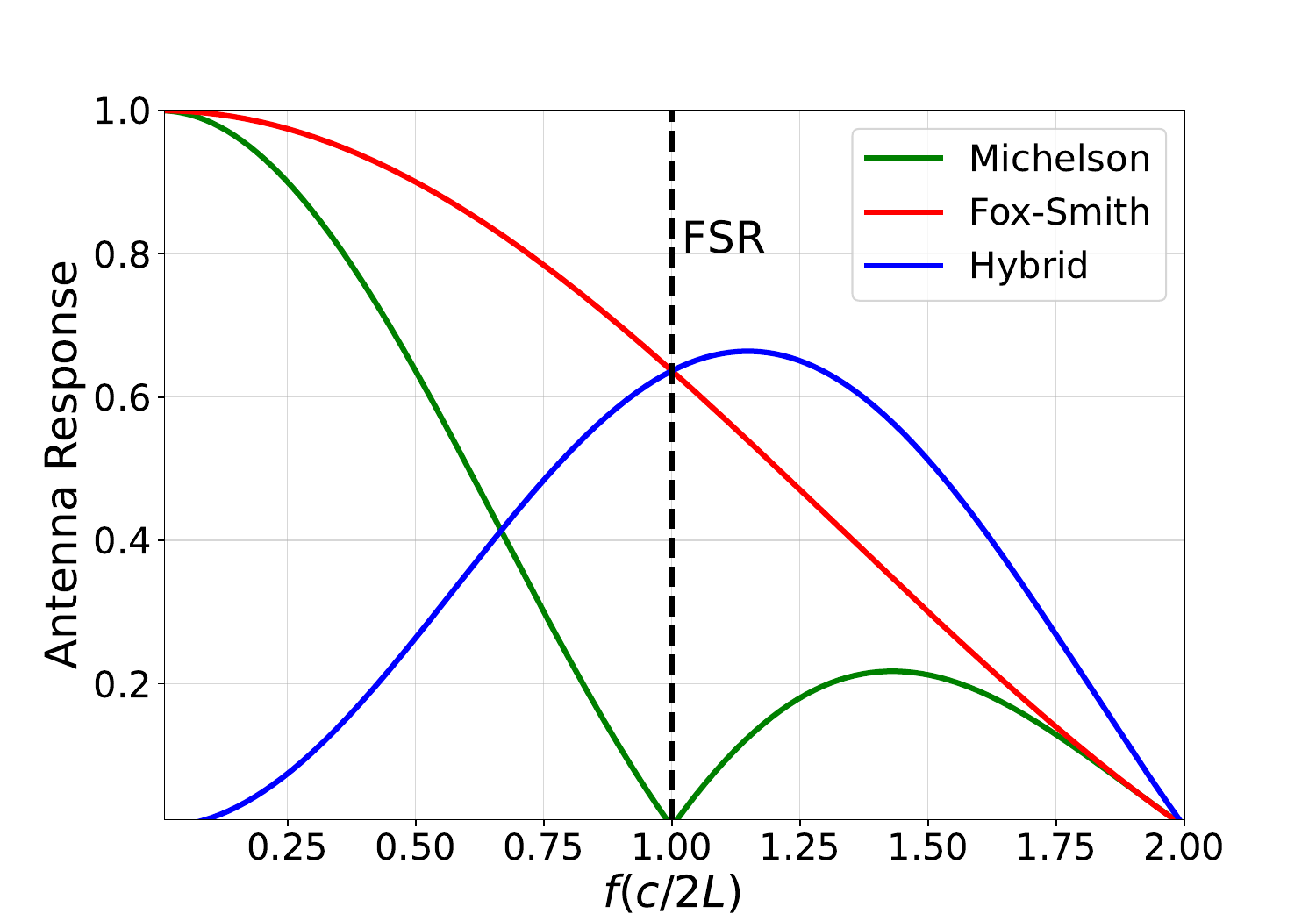}

    \end{minipage}
\caption{Comparison of the normalised antenna response to gravitational signals of three configurations, where the frequency is normalized to ${c}/{2L}$. Only the case of normal incidence for the plus polarisation is shown for illustration.}
\label{fig:antennares}
\end{figure}

Both employing L-shaped folded cavities as arm cavities, the subtle distinction between antenna response of configuration and hybrid configuration is primarily caused by their different optical coupling to the readout, as illustrated in Fig.\,\ref{fig:antenna}. In Fox-Smith configuration, the L-shaped cavity is utilized independently as a dual-output resonator, and light could emit after traversing half a round-trip within the cavity. As a comparison, in hybrid configuration, L-shaped cavities is used as 
arm cavities with single output, and light must propagate through multiple round-trips within the cavity before it transmits to the readout. Thus, the strain to displacement transfer function of Fox-Smith configuration is actually the phase-averaged response in half a round-trip,  and its responsive cavity length for GW signal is only $L/2$:
\begin{equation}
{\cal B}_{\rm fox-smith}(f) = \frac{L}{2}{\rm sinc} \left(\frac{\pi f L}{c}\right)\equiv \frac{L}{2}{F}_{\rm fox-smith}(f)\,.
\end{equation}
In contrast, the total round-trip of the field propagation in the hybrid configuration has four linear segments, and the transfer function from strain to displacement can be analytically expressed as:
\begin{equation}
\begin{aligned}
{\cal B}_{\rm hybrid} (f)& = \left\lvert \frac{1-e^{i\pi f L/c}}{\pi f/c}  (e^{-3i\pi f L/2c}-e^{-i\pi f L/2c}\right. \\
& \quad \quad \quad \quad \quad \quad \left. -e^{3i\pi f L/2c}+e^{3i\pi f L/2c}) \right\rvert \\ 
& = 2L \sin^2\left(\frac{\pi f L}{2c}\right)\,{F}_{\rm fox-smith}\,.
\end{aligned}
\end{equation}
Fig.\,\ref{fig:antennares} is a direct comparison of the antenna response of three configurations. At the first free spectral range, the hybrid configuration and Fox-Smith configuration maintain an identical high response to gravitational wave signals, which is significantly different from the null response of the Michelson configuration. This indicates that the sign-flipping effect caused by L-shaped cavity does not depend on the cavity's optical coupling to the output. The differences in antenna response between the hybrid configuration and Fox-Smith configuration due to optical coupling mainly locate in the low frequencies. The hybrid configuration have a null antenna response for gravitational wave signal at low frequencies, while the Fox-Smith configuration retain high antenna response to low-frequency signals due to the half-period leakage. In conclusion, from the perspective of antenna response, the hybrid configuration effectively inherits the great performance of the Fox-Smith configuration, at the cost of totally losing detecting sensitivity at low frequencies.

% \textcolor{red}{Optical response is defined as the optical amplification effect of the configuration on cavity strain signals. Specifically, for a configuration with arm cavity length $L$, the AC spectrum of electric field at the output under differential mode of cavity strain $h_{\rm cav}$could be expressed in a general form:}

% \textcolor{red}{\begin{equation}
% \widetilde E_{out}(f)=\frac{4\pi iL}{\lambda}e^{2\pi fL/c}\, \overline E_0  h_{\rm cav}(f)\,{\cal G}(f)=\widetilde E_{0}(f) \,{\cal G}(f).\,
% \end{equation}}
% \textcolor{red}{Here, $\lambda$ and $\overline E_0$ is the laser wavelength and the static field in the cavity respectively, $\widetilde E_{0}(f)$ represents the AC field spectrum resulting from  single reflection of mirror at distance $L$ under the same condition, and the dimensionless factor ${\cal G}(f)$ is the expression of optical response for configuration.}

Optical response is defined as the transfer function from mirror displacement to the output field normalized by the field spectrum of single reflection on the mirror. Both applying the optical structure of Michelson interferometer, traditional Michelson configuration and the hybrid configuration share identical optical responses, which significantly differ from the optical response of the Fox-Smith configuration that employs a dual readout arm cavity and has a completely different coupling to the readout. Assuming ETMs and CTMs are all ideal reflective mirrors, the optical response of three configurations could be then expressed as follows:

\begin{equation}
{\cal G}_{\rm mich} (f)=\frac{\sqrt{2}\,t_i}{1 - r_i e^{4i\pi f L/c}} \,,
\end{equation}

\begin{equation}
{\cal G}_{\rm fox-smith}(f)=\frac{\sqrt{2}t_i}{1 + r_i e^{2i\pi f L/c}} \,.
\end{equation}

\begin{equation}
{\cal G}_{\rm hybrid}(f) ={\cal G}_{\rm mich}(f)\,,
\end{equation}
where $r_i$ is the amplitude reflectivity of ITMs. The factor of $\sqrt{2}$ in all the three equations describes the final combination of two channels of differential mode signals by the beam splitter. 
 
We can see that near the first free spectral range $f=c/2L$, all three configurations exhibit high optical gain. Under the same ITM reflectivity, the peak optical response of the Michelson configuration and hybrid configuration is the same as that of the Fox-Smith configuration, while the effective bandwidth is half that of the Fox-Smith configuration. Conversely, under the same detection bandwidth, the peak optical response of the Michelson configuration is $1/\sqrt{2}$ times of that of the Fox-Smith configuration. The high optical gain near the first free spectral range ensures the hybrid configuration's sensitivity to high-frequency gravitational waves. Meanwhile, the Michelson and Fox-Smith configurations exhibit more pronounced differences in the optical response to low-frequency differential mode displacement. The Michelson configuration maintains high optical gain for low-frequency signals, whereas the Fox-Smith configuration exhibits optical suppression for low-frequency signals, significantly affecting the low-frequency sensitivity and sensing control strategy of the Fox-Smith configuration. 

\begin{figure}[htbp]
\centering

    \begin{minipage}[t]{0.95\linewidth}
        \centering
        \includegraphics[width=\textwidth]{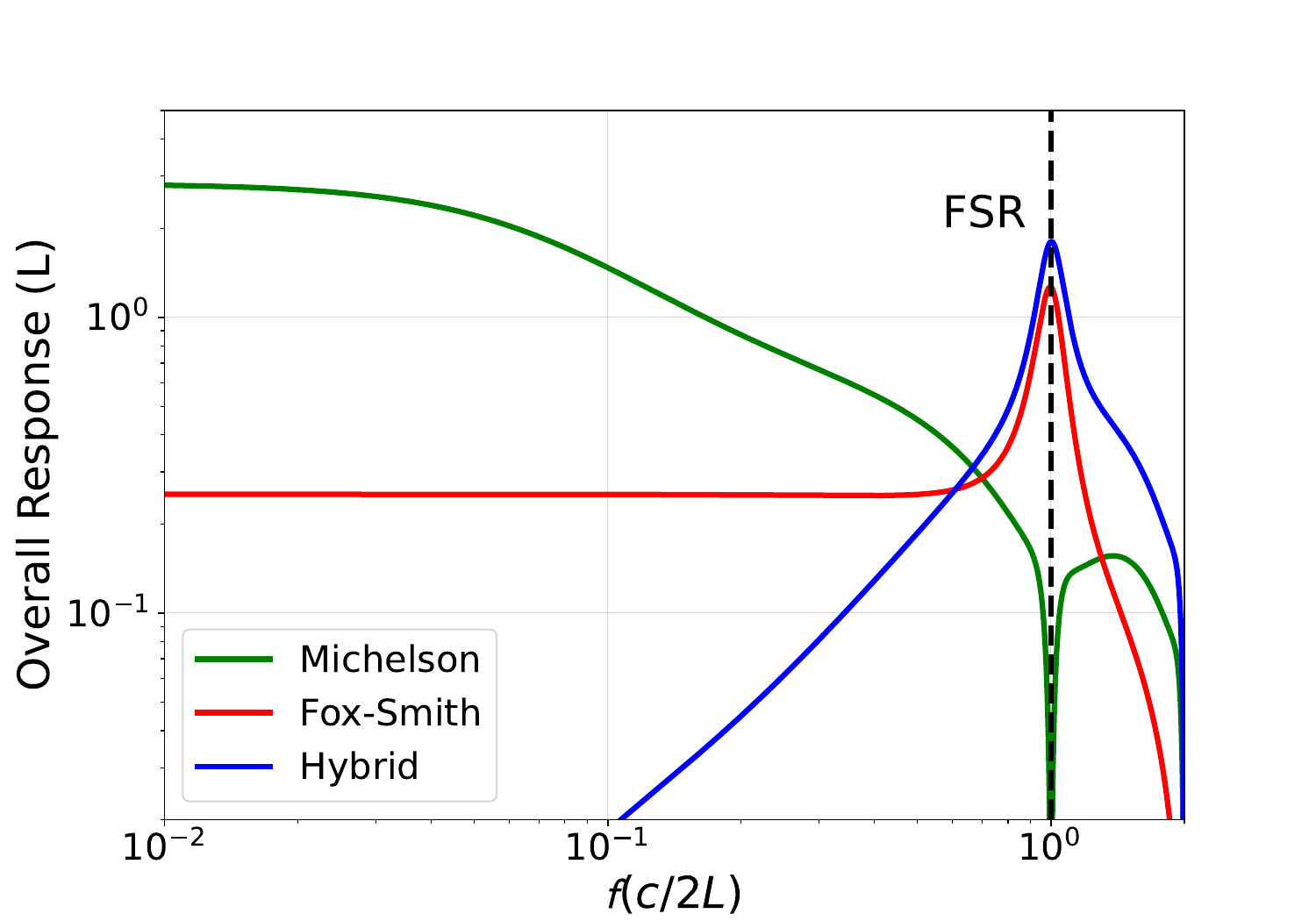}

    \end{minipage}
\caption{Comparison of overall response to gravitational signals of three configurations under the special case of $h_+$ polarization mode and normal incidence, with the same total cavity length, circulating power and optical bandwidth of the arm cavity. }
\label{fig:ant}
\end{figure}

The overall response of each configuration can be represented by the product of strain-to-displacement transfer function and optical response:

\begin{equation}
{\cal T}(f) = {\cal B}(f)\,{\cal G}(f)\,. 
\end{equation}
A comparison of the overall response to gravitational wave signal of three configurations is illustrated in Fig.\,\ref{fig:ant}. 
From the perspective of detecting capability, the hybrid configuration achieves high sensitivity to gravitational wave signals near the first free spectral range by combining the high-frequency specialized antenna response of the L-shaped cavity with the optical response of the traditional Michelson configuration. The peak sensitivity is $\sqrt{2}$ times higher compared to Fox-Smith configuration, a natural result of an additional arm cavity. As a trade-off, the hybrid configuration completely loses its ability to detect low-frequency gravitational wave signals due to its null antenna response at low frequencies. This indicates that the hybrid configuration is a detector design specialized for higher frequencies.

At the same time, from the perspective of longitudinal sensing and control, the hybrid configuration can directly adopt the control strategy of the Michelson since it shares the identical optical response to displacement as traditional Michelson configuration. Combining the high sensitivity response to high-frequency signals provided by L-shaped arm cavities with the well-studied optical behavior of the conventional Michelson configuration, the hybrid configuration emerges as a feasible scheme for kilohertz gravitational wave detection.

\section{ Noise Budget}\label{sec:opt}
In this section, we present the detailed noise budget of a conceptual detector with hybrid configuration. The enhanced performance of hybrid configuration in kilohertz frequency range is primarily governed by its unique antenna response. Meanwhile, the presence of a additional corner steering mirror in the L-shaped cavity also introduces specific differences to both the quantum and classical noise. 

In consistent with the intuitive working principle, the noise budget here is modeled under the specific case of normal incidence (incidence angle $\theta =0$). For gravitational waves incident from general directions, the noise level of the Hybrid design can be expressed by that of the normal incidence case as follows:

\begin{equation} \label{generalangle}
h_n^{\theta,\phi}(f) = \frac{F^{0,0}(f)}{F^{\theta,\phi}(f)} h_n^{0,0}(f)\,,
\end{equation}
where $F(f) = \sqrt{F_+^2+F_{\times}^2}$ is the magnitude of antenna response under specific direction of incidence. In \ref{sec:appendix B}, the antenna response of Hybrid configuration with arbitrary incidence angle is presented. By combining Eqn.\,\ref{generalangle} with the noise budget under normal incidence, the detection capability of Hybrid configuration for sources from different directions can be also evaluated. 

\subsection{Quantum Noise}

Firstly, we address the contribution of the corner  steering mirror to quantum noise, which constitutes the principal constraint on the sensitivity of hybrid configuration. To begin with, a basic overview of quantum noise of DRFPMI configuration is provided. The single-sided power spectral density of quantum noise then takes a simple form\cite{KLMTV}:
\begin{equation}
S_Q(\Omega)=\frac{4 \hbar}{M \Omega^2 L^2}\left(\frac{1}{\mathcal{K}}+\mathcal{K}\right)\,,
\end{equation}
where the first term represents shot noise, and the second term represents radiation pressure noise. Here, $\Omega=2\pi f$ is the signal frequency, $M$ is the mass of the end test mass, which is identical to that of the initial test mass, $L$ is the cavity length, and the dimensionless parameter $\cal{K}$ quantifies the intensity of conversion from amplitude noise to phase noise through the radiation pressure effect, which could be further expressed as:
\begin{equation}
{\cal K} = \frac{4  \omega_0 P_{\mathrm{arm}} t^2_{\mathrm{src}}}{c^2 M\left|e^{2 i \Omega \tau}-{r_{\mathrm{src}}}\right|^2 \Omega^2}\,,
\end{equation}
where $\tau={L}/{c}$ denotes the single-trip propagation time in the linear cavity, $\Omega_0$ is the laser frequency, $P_{\mathrm{arm}}$ is the laser power running in the cavity, and $t_{\mathrm{src}}$ and $r_{\mathrm{src}}$ are the amplitude transmissivity and reflectivity of the coupled cavity formed by signal recycling mirror and initial mirror.

% The quantum noise within traditional Michelson configuration is constrained by a minimum threshold known as the Standard Quantum Limit:
% \begin{equation}
% S_Q(\Omega)\ge S_{SQL}(\Omega)=\frac{8 \hbar}{M \Omega^2 L^2}\,,
% \end{equation}

In hybrid configuration, the  corner mirror amplifies radiation pressure effect. By treating end mirror and corner mirror as distinct sources of radiation pressure noise, we can analytically express the amplification effect for L-shaped folding on quantum noise of the arm cavity:
\begin{equation}
\mathcal{K}^* = \left[{1 + \frac{M}{M^*}\cos^2\left(\frac{\Omega \tau}{2}\right)} \right]\, \mathcal{K}\,,
\end{equation}

\begin{equation}
S_{Q}^*(\Omega)=\frac{4 \hbar}{M \Omega^2 L^2}\left(\frac{1}{\mathcal{K^*}}+\mathcal{K^*}\right)=\frac{4 \hbar \mathcal{K^*}}{ \mathcal{K} M \Omega^2 L^2}\left(\frac{1}{\mathcal{K^*}}+\mathcal{K^*}\right)\,,
\end{equation}

% \begin{equation}
% S_{SQL}(\Omega)=\frac{8 \hbar}{M \Omega^2 L^2}\left[{1 + 4\cos^2\left(\frac{\Omega \tau}{2}\right)} \right]\,.
% \end{equation}

Here, $M^*$ is the mass of the corner test mass. Further considering the antenna response of the hybrid configuration, this power spectrum can be transformed into the form of effective strain:
\begin{equation}
h_{Q}^*(\Omega)=\frac{\sqrt{S_{Q}^*(\Omega)}}{{F}_{\rm hybrid}(\Omega)},
\end{equation}

\begin{figure}[t!]
\centering

    \begin{minipage}[t]{\linewidth}
        \centering
        \includegraphics[width=\textwidth]{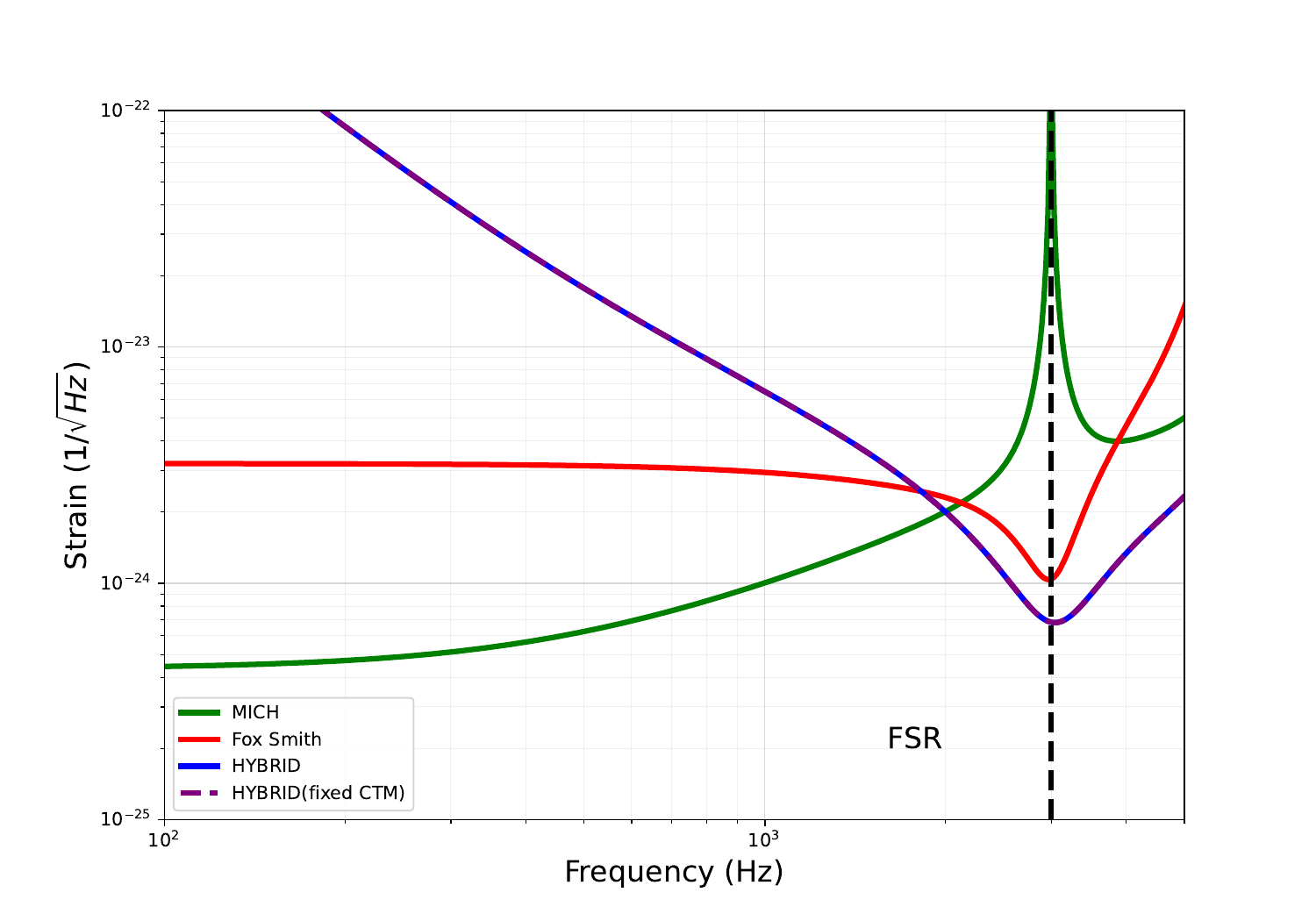}

    \end{minipage}
\caption{Quantum vacuum noise of the three configurations under the special case of normal incidence, with the same optical bandwidth under non-squeezing and lossless condition, where the optical length of signal recycling cavity is ignored. }
\label{fig:nosqz}
\end{figure}

Fig.\,\ref{fig:nosqz} illustrates a comparison of the quantum noise levels among the traditional Michelson, Fox Smith, and hybrid configurations without squeezing light injection and quantum losses. The hybrid configuration exhibits a sensitivity at the free spectral range approximately $\sqrt{2}$ times greater than that of Fox-Smith configuration, and comparable to the low-frequency detecting sensitivity of the traditional Michelson configuration (approximately $2/\pi$ times), aligning with conceptual analysis presented in Section \ref{sec:overview}. Additionally, given the hybrid configuration's operation within kilohertz range, where shot noise predominantly influences the primary quantum noise source, the corner mirror's induced additional radiation pressure does not notably affect the overall sensitivity of this configuration.

\begin{figure}[t!]
\centering

    \begin{minipage}[t]{\linewidth}
        \centering
        \includegraphics[width=\textwidth]{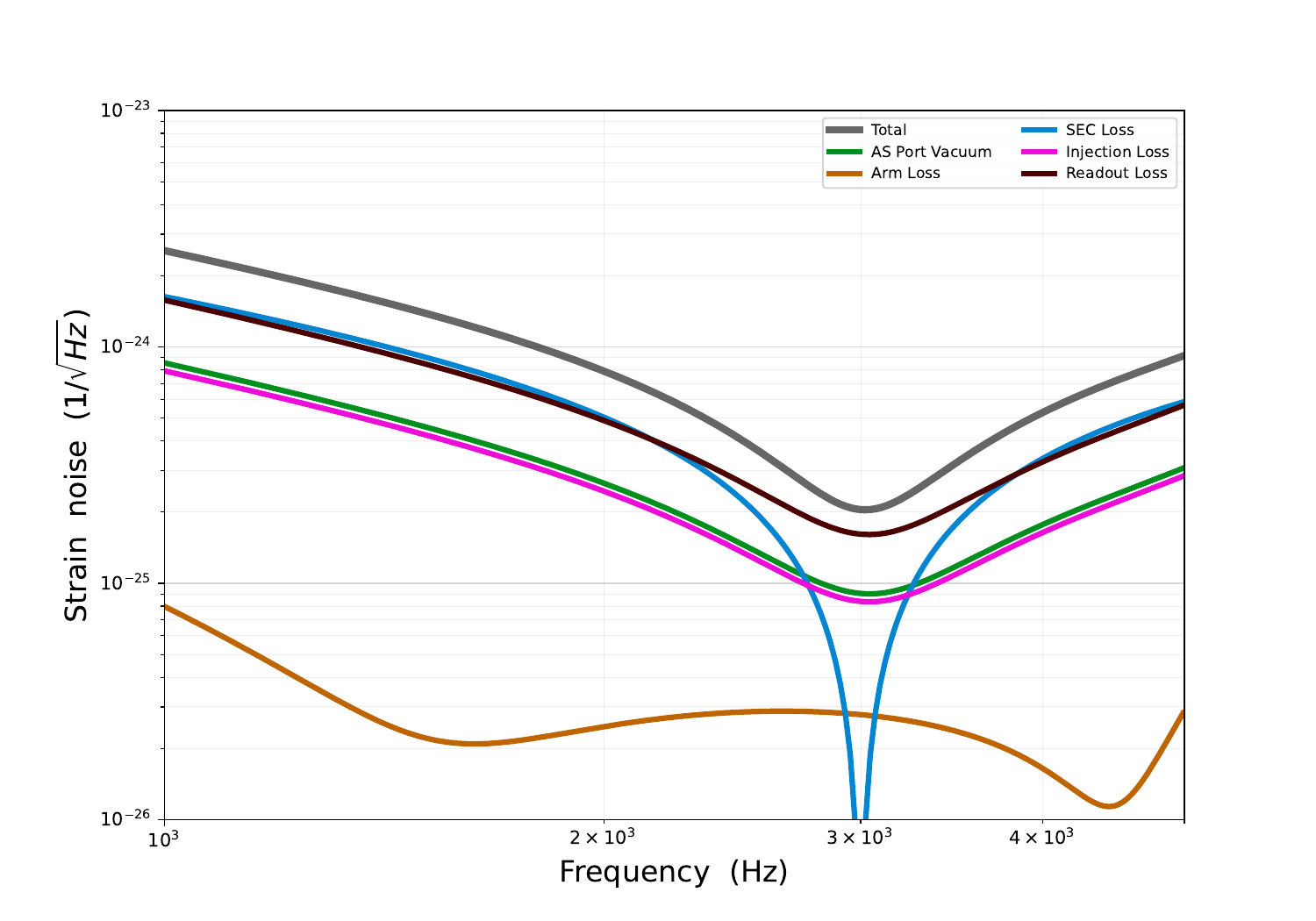}

    \end{minipage}
\caption{Quantum noise budget of the hybrid configuration at kilohertz range, with 11 dB of effective squeezing and quantum loss effects of auxiliary cavities. Detailed parameters are listed in Sec.\,\ref{sec:appendix A}.}
\label{fig:quantum}
\end{figure}

Quantum losses constitute a crucial aspect of quantum noise. In the context of hybrid configuration, the noise induced by quantum loss follows the same pattern as Michelson configuration. In Fig.\,\ref{fig:quantum}, comprehensive quantum noise budget for the configuration is presented. While modelling the quantum noise budget, we primarily adopted the squeezing level selection and quantum noise estimation methodology from Fox-Smith configuration.\cite{new.configuration} The only slight difference is that we choose the internal loss of single arm cavity as approximately 80 ppm. Under high squeezing level, quantum loss effects will supplant quantum shot noise as the predominant limitation on the sensitivity of hybrid configuration.

\subsection{Classical Noise}
Corner test masses also contribute non-quantum noise. Firstly, they act as independent sources of thermal noise.
 As light beam undergoes dual reflections on corner mirror during a single round-trip, the overall thermal noise emanating from CTM is represented as the coherent summation of the noise from these two reflections. Accounting for the $\sqrt{2}$ efficacy of the displacement of the corner mirror converted into cavity length, analytical expression for the contribution of the CTM to thermal noise is articulated as:
\begin{equation}
{S_{CTM}^{\,\, i}}={2}\cos^2\left(\frac{\Omega \tau}{2}\right)\,{S_{CTM0}^{\,\,i}}\,.
\end{equation}
Here, the upper index $i$ labels categories of thermal noises, while $S^{\,\, i}_{CTM0}$ signifies the single-sided power spectra of thermal noise for a mirror sharing identical coating and substrate parameters, injected laser power and beam size with CTM but serving as end mirror in the cavity.{For coating Brownian noise, the contribution of CTM should be further amplified by $3/2$ due to the excess in the power spectral density}\cite{foldedcavity}. 

It should be noted that while corner mirrors contribute to most types of thermal noise, they do not affect thermo-refractive noise due to their role as high-reflectivity mirrors within the cavity, where light does not traverse their substrate in the round-trip propagation. 

Apart from thermal noises, the presence of corner mirror also introduce certain corrections to other types of classical noises. For instance, vertical seismic and suspension thermal noise are scale-invariant but configuration-dependent. In the hybrid configuration, these noises ($\sqrt{S_{ext}^{\, i}}$) are $\sqrt{\frac{1}{2}+\frac{1}{2}\cos^2({\Omega \tau}/{2})}$ times compared to that of conventional Michelson configuration, where the first and second terms represent the contributions from the ETM and CTM, respectively. Other external noises scale as cavity length by $\sqrt{S_{ext}^{\, i}} \propto 1/L$. Among them, horizontal seismic\cite{seismic,seismic_2} and suspension thermal noise, Newtonian noise\cite{newtonian}, and the damping noise of residual gas\cite{gas_damping} can be considered as local noises on mirrors and would thus be scaled up by $\sqrt{1+2\cos^2({\Omega \tau}/{2})}$ compared to traditional configuration with the same arm cavity length. Scattering noise of residual gas\cite{gas_scattering}, however, only depends on the total distance that beam travels, thus remaining unchanged between these two configurations.

\subsection{Total Noise Budget}

Total noise budget of the conceptual detector in hybrid configuration with each arm spanning 25 km is numerically modeled by adding all the correlations above into the noise-calculating software, pyGWINC\cite{GWINC}. In our modeling approach, we utilize a laser wavelength of 1064 nm and mirror material of fused silica, identical to those used in Advanced LIGO.\cite{Aasi2015c}. Like other third-generation detectors\cite{CE20,CE40,Voyager}, we anticipate high circulating power in arm cavities, thus setting the static running power in each arm cavity at 1.5 MW. To ensure an adequate detection bandwidth, we opt for a relatively large ITM transmissivity of $T_{ITM}= 0.014$ . We also apply a high-finesse signal recycling with a signal recycling mirror transmissivity of $T_s = 0.03$, further broadening the effective detection bandwidth to approximately 500 Hz.

As for the parameters of L-shaped cavity, we directly implement the original design of detector under Fox Smith configuration\cite{new.configuration}. Specifically, radius of curvature of two ITMs and ETMs are set as 40 km, while CTMs are designed to be flat. This choice results in a waist size of $w_0 = 8.0$ cm and a beam size of $w_I = w_E = 13.2$ cm at ITMs and ETMs, respectively. Meanwhile, due to the $45^{\circ}$ tilt of the corner mirror, the spot on CTMs are ellipses with a semi-major axis of $\sqrt{2}w_0$ and a semi-minor axis of $w_0$. All mirrors in the arm cavities are selected with a radius of $34$ cm, while the thickness of CTMs/ETMs(ITMs) are set as $17/40$cm , resulting in a total mass of 136 kg/320 kg respectively.

As a conservative estimation, we employ parameters from second-generation gravitational wave detectors in our modelling of classical noise. We adopt the Ta2O5-SiO2 bi-layer coatings\cite{coating} and fused silica substrate\cite{Aasi2015c} used in Advanced LIGO. For the passive damping system, we utilize the same quadruple pendulum suspension system as LIGO\cite{LIGO_suspension,CE_suspension}, on the ETMs and CTMs. The vacuum requirements for the arm cavity align with those of Advanced LIGO.

\begin{figure}[htbp]
\centering

    \begin{minipage}[t]{\linewidth}
        \centering
        \includegraphics[width=\textwidth]{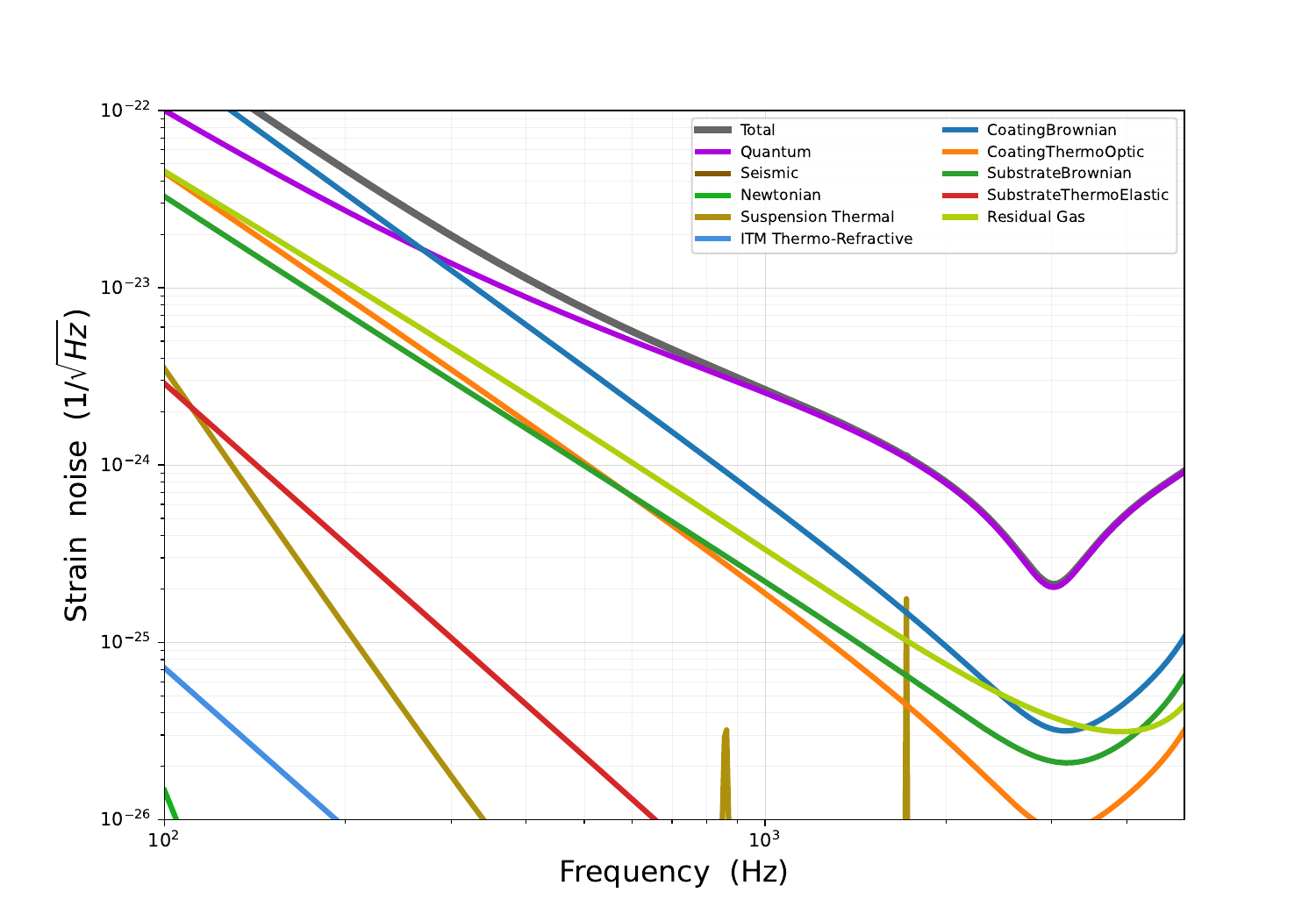}

    \end{minipage}
\caption{Overall noise budget of the new detector in hybrid configuration. For the gravitational wave signals in the frequency range of 2-4 kHz, corresponding to the merger process of binary neutron stars, the hybrid configuration exhibits higher detection sensitivity compared to Fox Smith configuration. The noise budget is modelled by the simulation package, pyGWINC\cite{GWINC}.}
\label{fig:totnoise}
\end{figure}

Total noise budget of the conceptual detector is illustrated in Fig.\,\ref{fig:totnoise}. The result indicates that hybrid configuration exhibits higher sensitivity to gravitational wave detection in the 2-4 kHz range compared to the new configuration, especially in the interval above the first Free Spectral Range(3000 Hz). This highlights the potential and advantages of the hybrid configuration as a high-frequency gravitational wave detector. Meanwhile, as the new configuration operates within kilohertz frequency band, quantum noise emerges as its principal noise source, whereas classical effects such as seismic disturbances, suspension dynamics, and thermal influences do not impose fundamental constraints on its detection sensitivity even under the technical constraints akin to second-generation detectors. The insensitivity to materials and active or passive noise filtering further enhance the feasibility of hybrid configuration.

\section{Conclusion and outlook}\label{sec:conclusion}
Kilohertz gravitational waves, instrumental in studying the late inspiral and merger of binary neutron stars, hold paramount significance in astrophysics. However, current kilohertz gravitational wave detectors, which primarily rely on the traditional Michelson configuration, face limitations induced by the antenna response of linear Fabry-Perot arms. The antenna pattern experience a cancellation at FSR, which makes it not the optimal choice for detecting high frequency signals near or exceeding the FSR.

Our study combines L-shaped resonator, A 90-degree angle folded cavity that optimizes the response to gravitational waves near the FSR, with traditional Michelson configuration. The merged design basically retains the high-frequency sensitivity of the former and the well-established sensing and control scheme of the latter. Conceptually, the hybrid configuration demonstrates superior detection sensitivity in 2-4 kHz range compared to existing configurations, underscoring its potential to advance kilohertz gravitational wave astronomy. Further development of hybrid configuration will primarily focus on technical issues, including alignment control of the steering corner mirror, optimal mode matching in coupled cavities, and the laser noise suppression around the first cavity FSR.

Meanwhile, as one of the pioneering works employing L-shaped folded cavities, this study provides a detailed discussion on the effects and principles of antenna response changes induced by cavity folding. Designing the shape of the arm cavity to optimize the detection sensitivity of the detector is also a feasible and prospective direction for future design of high-frequency gravitational wave detector.

\section{Acknowledgements}
We would like to acknowledge the support from the 'Gravitational Wave Detection' program (2023YFC2205800 and 123B1015) funded by the National Natural Science Foundation of China, and support from Quantum Information Frontier Science Center. Also, we really appreciate Chunnong Zhao, Matthew Evans, Mikhail Korobko and AIC group for fruitful
discussions. X. G. and H. M. are supported by State Key Laboratory
of Low Dimensional Quantum Physics and the start-up
fund from Tsinghua University.  T. Z., D. M. acknowledge
the support of the Institute for Gravitational Wave Astronomy at the University of Birmingham, STFC Quantum Technology for Fundamental Physics scheme (Grant
No. ST/T006609/1), and EPSRC New Horizon Scheme
(Grant No. EP/V048872/1 ). T. Z. acknowledges the
support of department of Gravitational Waves and Fun-
damental Physics in Maastricht University and ETEST
project. D. M. is supported by
the 2021 Philip Leverhulme Prize.
\newpage
\bibliographystyle{unsrt}
\bibliography{apssamp}% Produces the bibliography via BibTeX.

\appendix
\renewcommand\appendix{\par
    \setcounter{section}{0}
    \setcounter{subsection}{0}
    \gdef\thesection{ \Alph{section}}}
    
\section{Parameters }\label{sec:appendix A}
The detailed parameters of the conceptual detector with Hybrid configuration is illustrated in Table.\,\ref{tab:table1}.
\begin{table}[b]
\caption{\label{tab:table1}
Parameters of the 25km Hybrid detector}
\begin{ruledtabular}
\begin{center}
\begin{tabular}{ccc}
Description& Value& Units\\
\hline
Power transmission of input mirror& 0.014 &\\
Power transmission of SEM& 0.03 &\\
total cavity length & 50000& m\\
Signal extraction cavity length& 50.00& m\\
\hline
Input laser power& 250& W\\
Power inside power recycling cavity& 11 & kW\\
Power inside the L-shaped arm cavity& 1.5 & MW\\
laser wavelength& 1064 & nm\\
\hline
Arm loss & $80$ & $\mathrm{ppm}$ \\
SEC loss & $500$ & $\mathrm{ppm}$ \\
Input loss & $1.5$ & $\%$ \\
Output loss & $3.5$ & $\%$ \\
Effective squeezing level(for shot noise)& 11& dB \\
\hline
CTM mass & 136 & $\mathrm{kg}$  \\
ETM/ITM mass &  320 & $\mathrm{kg}$\\
ITM/ETM radius & 34 & $\mathrm{cm}$ \\
CTM radius &  34 & $\mathrm{cm}$ \\
ITM/ETM RoC & 40 & $\mathrm{km}$ \\
CTM RoC &  Inf &  \\
ITM/ETM beam size & 13.2 & $\mathrm{cm}$ \\
waist size of beam& 8 & $\mathrm{cm}$ \\
\hline
Substrate material & fused silica\\
loss angle of high-n/low-n coating & 3.89e-4/2.3e-5\\

\end{tabular}
\end{center}
\end{ruledtabular}
\end{table}
\par
\par

\section{Directional Dependence of the Antenna Response of Hybrid Configuration}\label{sec:appendix B}

In Section \ref{sec:overview}, we investigate the frequency-dependent response of the L-shaped Michelson configuration to $h_+$-polarized gravitational waves in a special case. In practical scenarios, gravitational wave sources exhibit spatial distributions across various directions, necessitating an examination of a configuration's response to gravitational wave signals across space to assess the feasibility of novel configurations. 

In this section, we derive the exact formula for the antenna response of the L-shaped Michelson configuration to gravitational wave signals incident from any direction. This formula offers a visual representation of the spatial distribution of sensitivity inherent to the new configuration. 

It is well-known that gravitational wave signals can be decomposed into polarization modes. The strain caused by gravitational wave propagating along the direction $\hat{n}$ could be expressed as:

\setcounter{equation}{0}
\renewcommand{\theequation}{B.\arabic{equation}}
\begin{equation}
h_{i j}(t, \vec{x})=h_{+}(t, \vec{x}) \varepsilon_{i j}^{+}(\hat{n})+h_{\times}(t, \vec{x}) \varepsilon_{i j(\hat{n})}^{\times}\,,
\end{equation}
where $\varepsilon_{i j}^{+}(\hat{n})$ and $\varepsilon_{i j}^{\times}(\hat{n})$ are normalized polarization tensor accounts for $h_+$ and $h_{\times}$ modes.

For a balanced interferometric gravitational wave detector, the final output is directly proportional to the differential relative change in the lengths of its two arm cavities, denoted as $\delta V \equiv (L_1-L_2)/L$, where $L_1$ and $L_2$ represent the lengths of each arm cavity, and $L$ is the average arm cavity length. Additionally, $\delta V$ can be decomposed into contributions from the two distinct polarization states of gravitational waves\cite{Essick_PRD17}:
\begin{equation}
\delta V=\frac{\delta L_x-\delta L_y}{L}=F_{+}(\hat{n}) h_{+}(t)+F_{\times}(\hat{n}) h_{\times}(t)\,.
\end{equation}

Here, the configuration-dependent scalar factors $F_{+}(\hat{n})$ and $F_{\times}(\hat{n})$, also known as the detector's antenna response, intuitively depict the efficiency with which gravitational wave signals of two polarization are converted into the detector's strain on differential degrees of freedom. Thus, the total magnitude of antenna response, $\sqrt{F_+^2+F_{\times}^2}$, serves as a valid indicator of the configuration's sensitivity in detecting gravitational waves. Furthermore, the antenna response factor can be decomposed into the contraction of a configuration-dependent tensor $D_{ij}$ and the polarization tensor as follows:
\begin{equation}
F_{+,\times}=D_{ij}\epsilon_{+,\times}^{ij}\,,
\end{equation}

where $D_{ij}$ could be derived straightforward by considering the round-trip travel time in each arm cavity along null geodesic\cite{Essick_PRD17,Rakhmanov_CQG08}. As the gravitational signal is a tiny perturbation, it can be effectively modeled as a monochromatic plane wave for analytical purposes. Assuming it has a general form of $h_{+, \times}(t , \vec{x})=h_{0+, \times}e^{i(\vec{k} \cdot \vec{x} -\omega t)}$, we could clearly see the relationship below:

\begin{equation}
h_{+, \times}(t , \vec{x})=h_{+,\times}\left(t-\frac{\hat{n} \cdot \vec{x}}{c}\right)\,.
\end{equation}

The total round-trip travel time of a photon within the arm cavity can be expressed as the summation of the travel times associated with its consecutive straight-line propagation segments. Considering a photon launched in the direction $\hat{a}$ and travelling for a total distance of $l$, the total propagation time along such geodesic could be expressed as follows:
\begin{equation}
c\left(t-t_0\right)=\int_0^{l}\left(1+h_{i j} a^i a^j\right)^{\frac{1}{2}} \cdot d \xi^{\prime}\,.
\end{equation}

By employing a linear expansion, the additional propagation time induced by gravitational waves can be derived:
\begin{equation}
\delta T=\frac{1}{2 c} a^i a^j \int_0^l h_{ij}\left(t_0+\frac{\xi}{c}-\frac{\hat{n} \cdot \hat{a}}{c} \xi\right) d \xi\,.
\end{equation}

The complete calculation of the additional time accumulated during a full round-trip travel directly provides the analytical expression for $D_{ij}$. Specifically, for folded Michelson configuration, the propagation direction of beam would be flipped each quarter of period in a single round-trip. In one of the arm cavities, the beam propagates along $\pm x$ axis in the first and last quarter of period, and propagates along $\pm y$ axis in the rest half period. Applying the expression of extra propagation time along null geodesic, we could get the extra time of the first and last quarter of round-trip:

\begin{equation}
\delta T_{11}=\frac{L}{4 c}\left(e_x\right)^i\left(e_x\right)^j \cdot h_{i j} \cdot e^{-i \pi f T_{x-}} \cdot \sin c\left(\pi f T_{x-}\right)\,,
\end{equation}

\begin{equation}
\delta T_{14}=\frac{L}{4 c}\left(e_x\right)^i\left(e_x\right)^j \cdot h_{i j} \cdot e^{-i \pi f (8T-T_{x+})} \cdot \sin c\left(\pi f T_{x+}\right)\,.
\end{equation}

where $T_{x\pm}=L(1\pm \frac{\hat{n} \cdot \hat{e_x}}{c})/2c$. Similarly, we could define $T_{y\pm}=L(1\pm \frac{\hat{n} \cdot \hat{e_y}}{c})/c$, then the gravitational-caused time delay in the second and third quarter could be expressed as:

\begin{align}
&\delta T_{12}+\delta T_{13} = \frac{L}{4 c} \left(e_y\right)^i \left(e_y\right)^j h_{i j} \, e^{-2i \pi f (T+T_{x-})} \notag \\
&\times \left[ e^{-i \pi f T_{y-}} \sin c\left(\pi f T_{y+}\right) 
+ e^{i \pi f T_{y+}} \sin c\left(\pi f T_{y-}\right) \right] \,.
\end{align}

Summing the extra time in the four quarters, we could get the analytical expression of the total extra time in a full round-trip in a single L-shaped arm cavity:
\begin{equation}
\begin{aligned}
\delta T_1 &= \frac{L}{4c} \, h_{ij} \Bigg[
    \left\{
        e^{-i \pi f T_{x-}} \sin c\left(\pi f T_{x-}\right) 
         \right.  \\
&\quad \left.+ e^{-i \pi f(8 T - T_{x+})}  \sin c\left(\pi f T_{x+}\right)
    \right\} \left(e_x\right)^i \left(e_x\right)^j  \\
&\quad + e^{-2 \pi i f(T + T_{x-})} 
    \left\{
        e^{i \pi f T_{y+}} \sin c\left(\pi f T_{y-}\right) \right.  \\
&\quad \left. + e^{-i \pi f T_{y-}} \sin c\left(\pi f T_{y+}\right)
    \right\} \left(e_y\right)^i \left(e_y\right)^j 
\Bigg]  \\
&= \frac{L}{4} \, h_{ij} \left[
    D_{1 x}(\hat{n},f)  (e_x)^i (e_x)^j 
    + D_{1 y}(\hat{n},f)  (e_y)^i (e_y)^j
\right] \,.
\end{aligned}
\end{equation}

The comprehensive formulation for the additional time in the opposing arm cavity could be acquired through a straightforward interchange of the directional vectors $\hat{e_x}$ and $\hat{e_y}$:
\begin{equation}
\begin{aligned}
\delta T_2 &= \frac{L}{4c} \, h_{ij} \Bigg[
    \left\{
        e^{-i \pi f T_{y-}} \sin c\left(\pi f T_{y-}\right) 
         \right.  \\
&\quad \left. + e^{-i \pi f (8 T - T_{y+})} \sin c\left(\pi f T_{y+}\right)
    \right\} \left(e_y\right)^i \left(e_y\right)^j  \\
&\quad + e^{-2 \pi i f (T + T_{y-})} 
    \left\{
        e^{i \pi f T_{x+}} \sin c\left(\pi f T_{x-}\right) \right. \\
&\quad \left. + e^{-i \pi f T_{x-}} \sin c\left(\pi f T_{x+}\right)
    \right\} \left(e_x\right)^i \left(e_x\right)^j 
\Bigg] \\
&= \frac{L}{4} \, h_{ij} \left[
    D_{2 x}(\hat{n},f)  (e_x)^i (e_x)^j 
    + D_{2 y}(\hat{n},f)  (e_y)^i (e_y)^j
\right] \,.
\end{aligned}
\end{equation}

The expected readout of the interferometer is proportional to the difference of extra time in the two L-shaped cavities: 
\begin{align}
\delta T &= \delta T_{1} - \delta T_{2} \notag \\
&= \frac{L}{4c} \, h_{ij} \Big[ 
    \big( D_{1 x}(\hat{n},f) - D_{2 x}(\hat{n},f) \big) (e_x)^i (e_x)^j \notag \\
&\quad + \big( D_{1 y}(\hat{n},f) - D_{2 y}(\hat{n},f) \big) (e_y)^i (e_y)^j 
\Big] \,.
\end{align}

Taking into account that the total optical path length for a round-trip is $2L$, the expression of $D^{ij}$ can be straightforwardly deduced from the aforementioned equation:

\begin{align}
D^{ij}(\hat{n},f) &= \frac{1}{8} \Big[
    \big( D_{1 x}(\hat{n},f) - D_{2 x}(\hat{n},f) \big) (e_x)^i (e_x)^j \notag \\
&\quad + \big( D_{1 y}(\hat{n},f) - D_{2 y}(\hat{n},f) \big) (e_y)^i (e_y)^j 
\Big] \,.
\end{align}

Additionally, when the polarization angle $\psi$ is set to 0, the polarization tensor $\epsilon_{+,\times}^{ij}$ of gravitational waves arriving from the direction ($\theta$, $\phi$) can be systematically expanded in the detector frame. This expansion furnishes the precise formulation for the antenna response factor:

\begin{equation}
\begin{aligned}
F_{+}=&\frac{1}{8}\Biggl\{ \Bigl[ D_{1x}(f, \hat{n})-D_{2x}(f, \hat{n}) \Bigr] \cdot \Bigl[ -\sin^2 \phi + \cos^2\theta \cos^2 \phi \Bigr] \\
&+ \Bigl[ D_{1y}(f, \hat{n})-D_{2y}(f, \hat{n}) \Bigr] \cdot \Bigl[ -\cos^2 \phi + \cos^2\theta \sin^2 \phi \Bigr] \Biggr\}\,,
\end{aligned}
\end{equation}

\begin{align}
F_{\times} &= -\frac{1}{8} \Big[ 
    D_{1 x}(f, \hat{n}) - D_{2 x}(f, \hat{n}) \notag \\
&\quad - D_{1 y}(f, \hat{n}) + D_{2 y}(f, \hat{n}) 
\Big] \cos \theta \sin 2\phi \,.
\end{align}

\begin{figure}[htbp]
\centering

    \begin{minipage}[t]{0.492\linewidth}
        \centering
        \includegraphics[width=\textwidth]{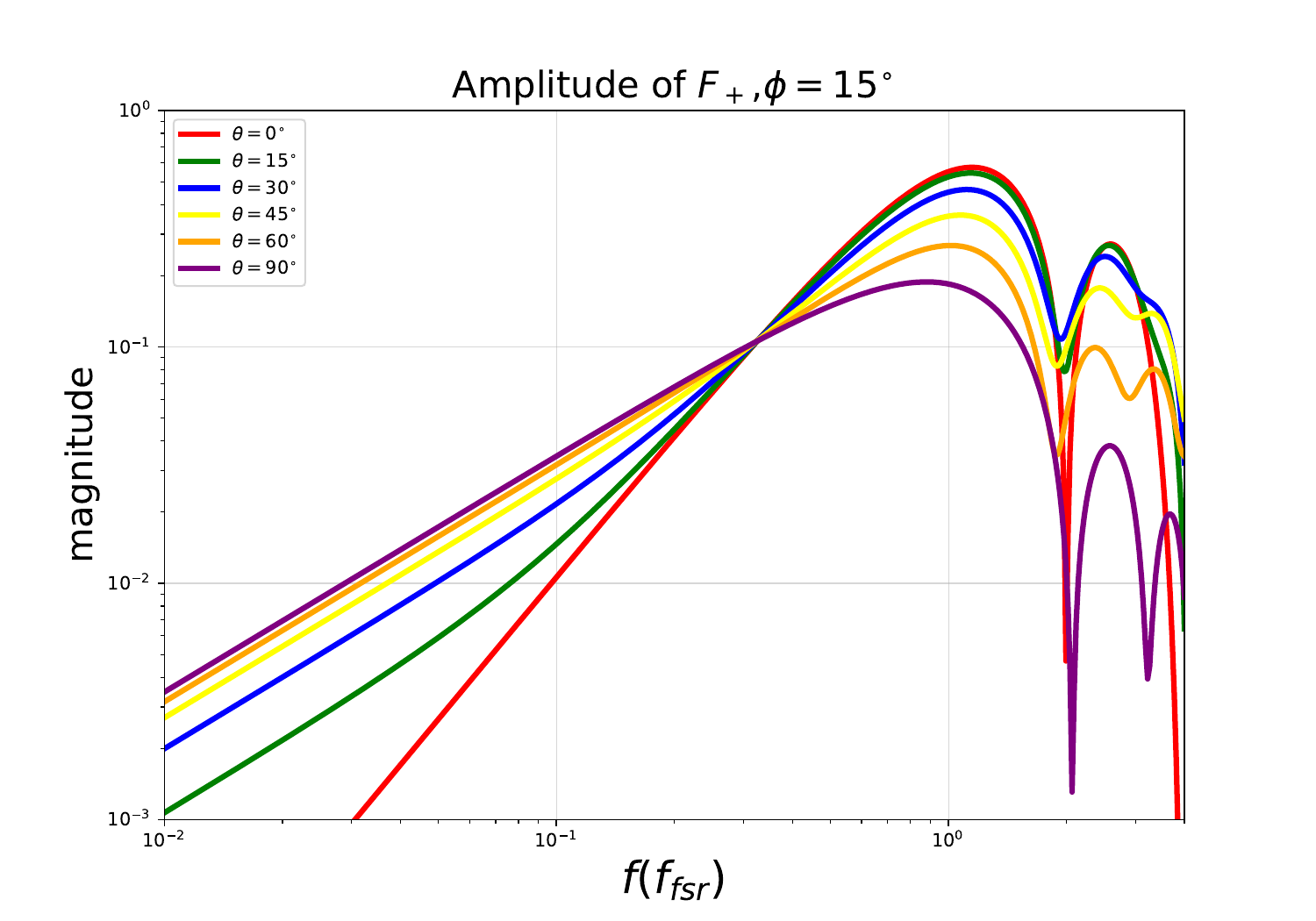}

    \end{minipage}
    \begin{minipage}[t]{0.492\linewidth}
        \centering
        \includegraphics[width=\textwidth]{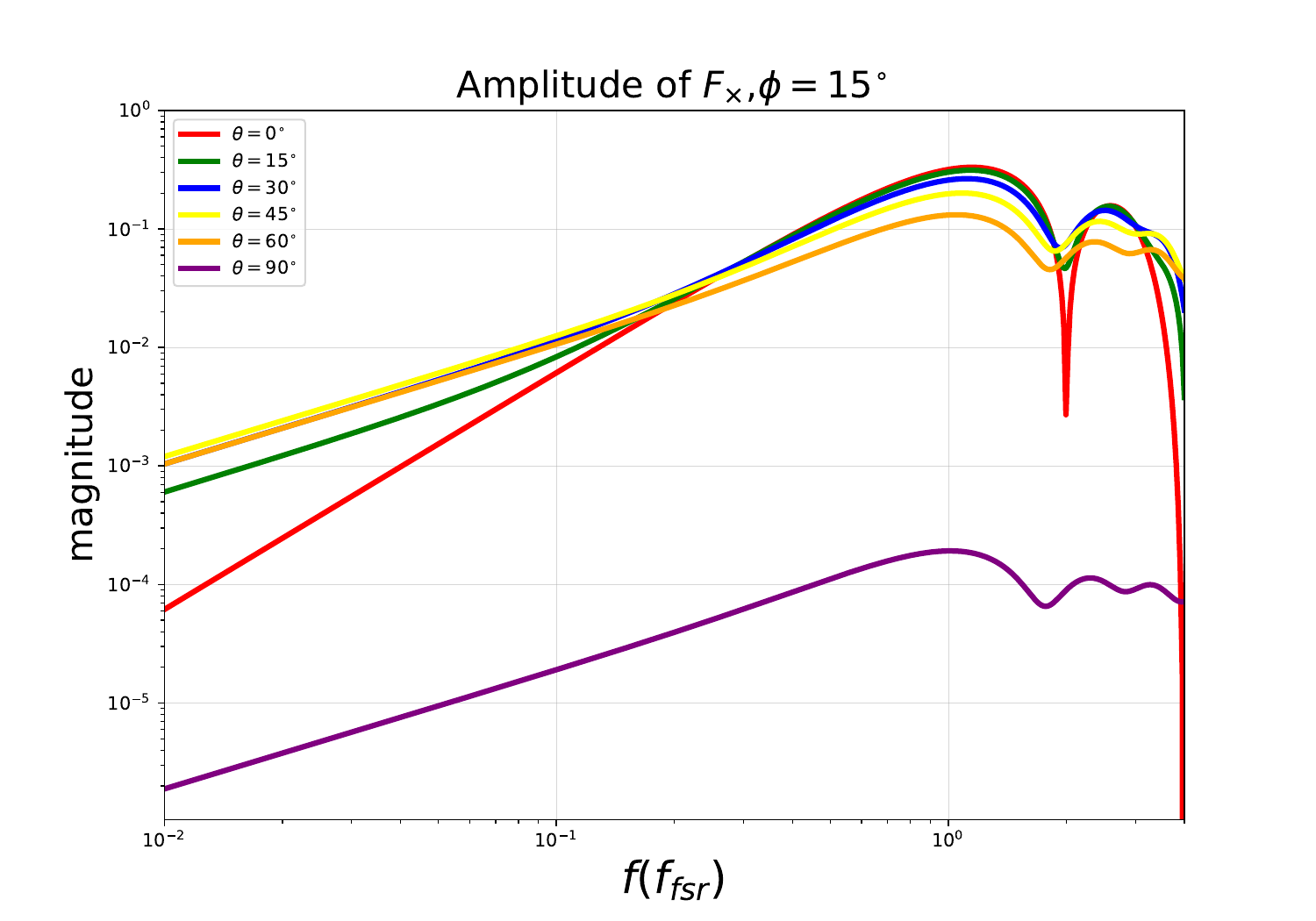}

    \end{minipage}
    \begin{minipage}[t]{0.492\linewidth}
        \centering
        \includegraphics[width=\textwidth]{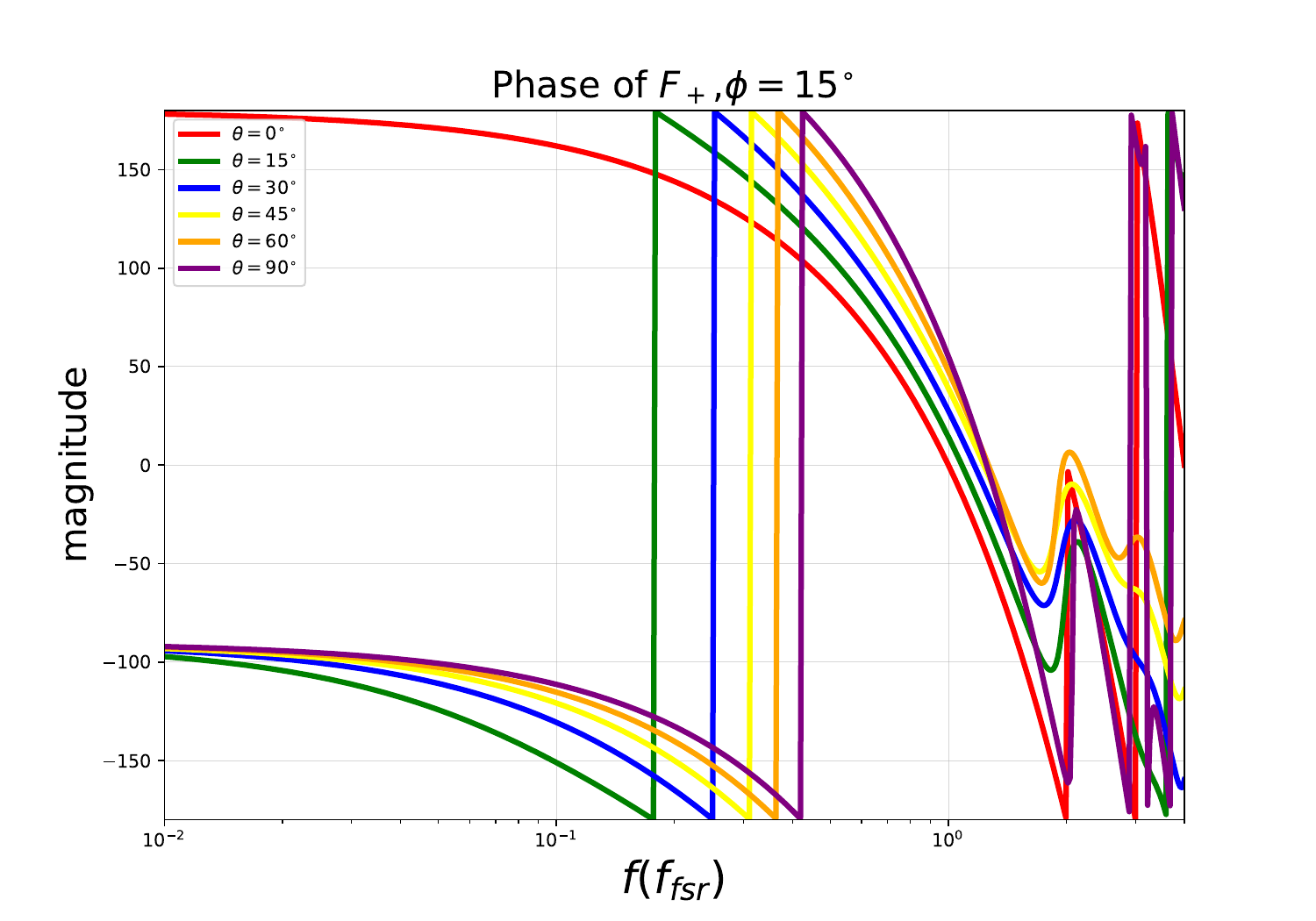}

    \end{minipage}
    \begin{minipage}[t]{0.492\linewidth}
        \centering
        \includegraphics[width=\textwidth]{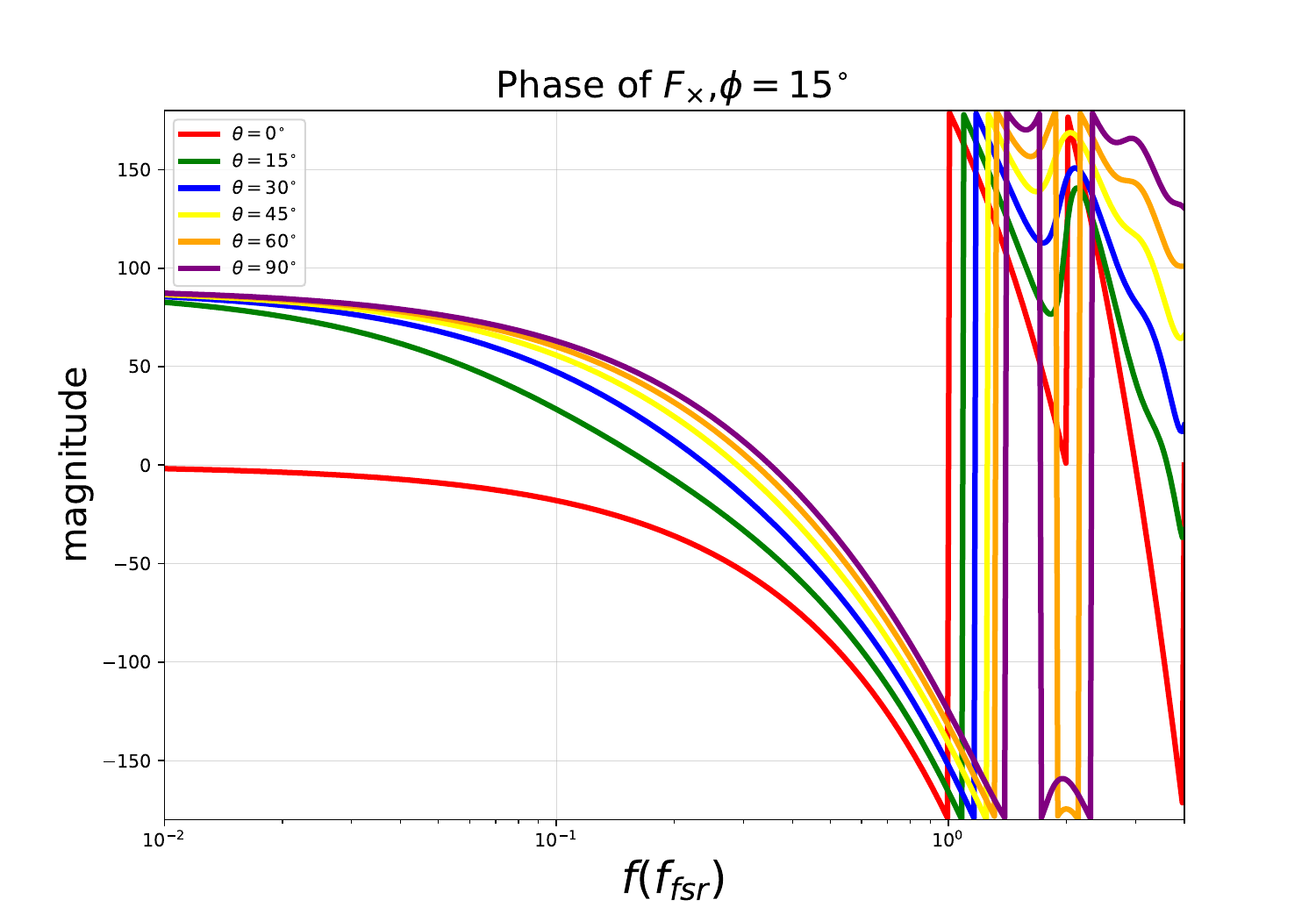}

    \end{minipage}
\caption{Bode Plots of antenna response $F_+$ and $F_{\times}$ at a few source directions $(\theta,\phi)$. We could note that the peak of antenna response is reached around $f=1.2f_{fsr}$. Considering the resonant condition of arm cavity, the peak detecting sensitivity is still reached at $f=f_{fsr}$ Meanwhile, the antenna response factor exhibits a swift convergence to zero at low frequencies, indicative of the system's null response to low-frequency gravitational waves. This inherent trait delineates the Folded Michelson configuration's capability solely for targeted detection of high-frequency gravitational waves.}
\label{fig:freq}
\end{figure}

\begin{figure}[htbp]
\centering

    \begin{minipage}[t]{0.492\linewidth}
        \centering
        \includegraphics[width=\textwidth]{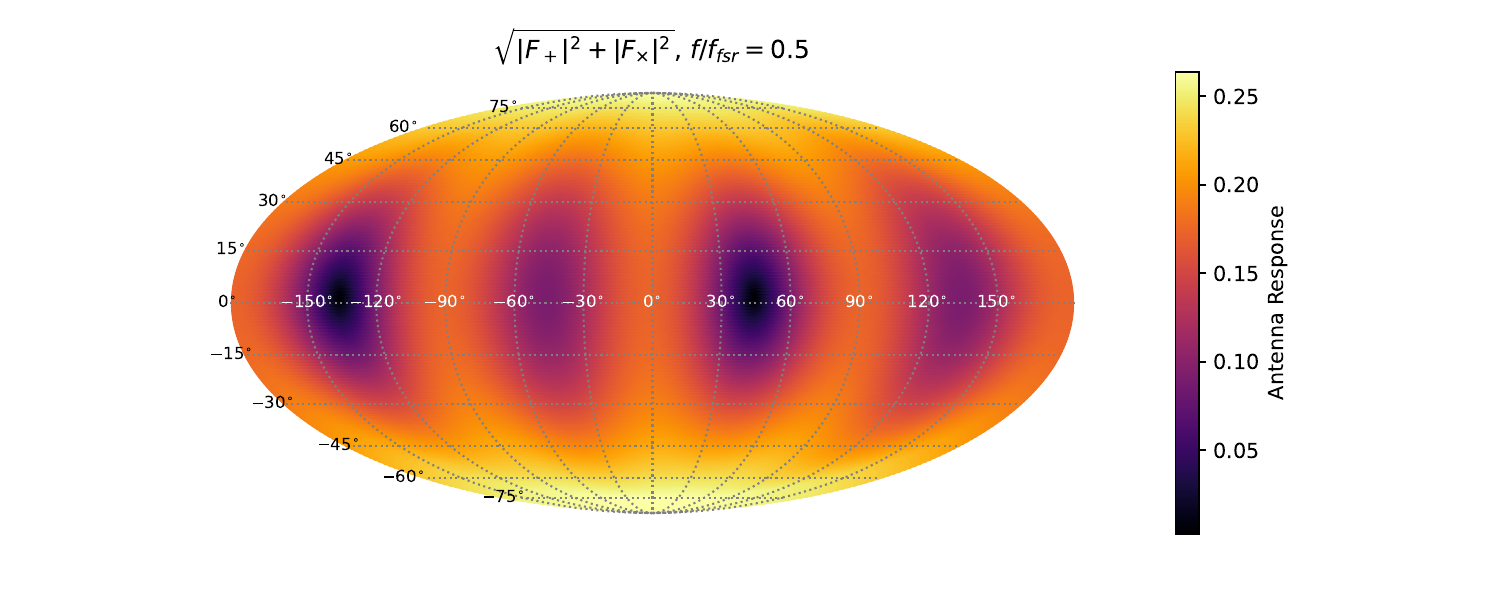}

    \end{minipage}
    \begin{minipage}[t]{0.492\linewidth}
        \centering
        \includegraphics[width=\textwidth]{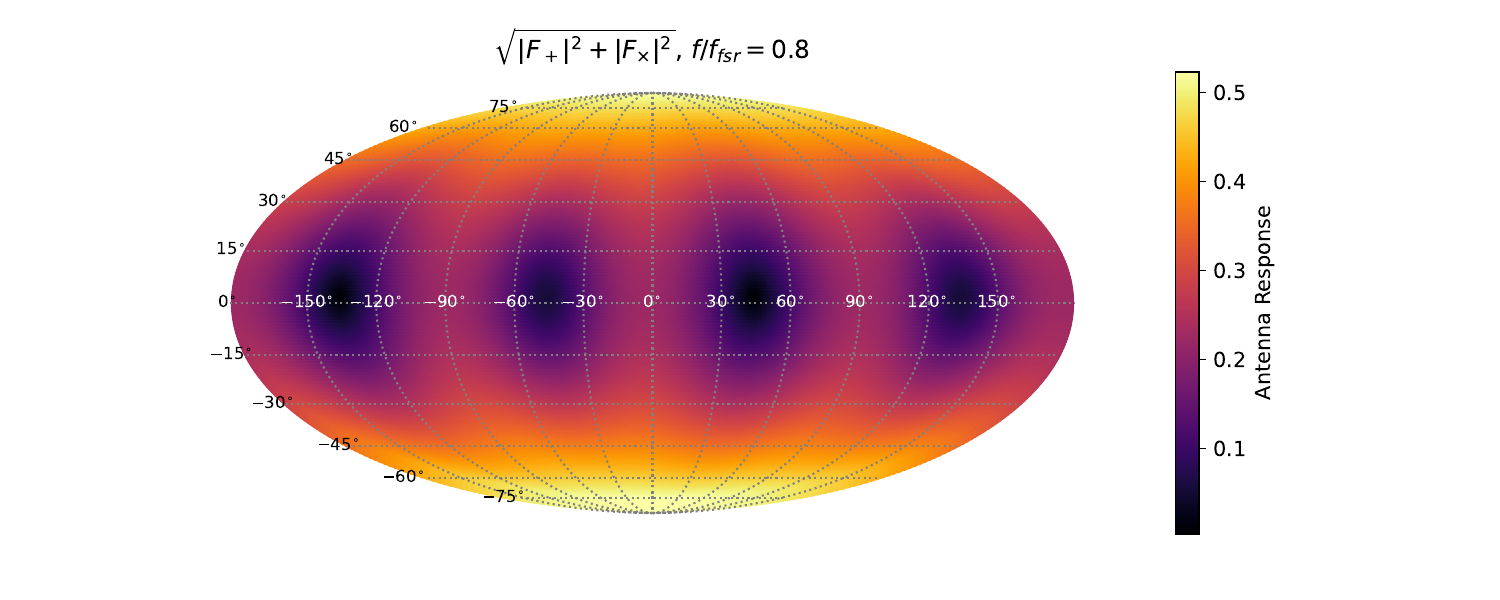}

    \end{minipage}
    \begin{minipage}[t]{0.492\linewidth}
        \centering
        \includegraphics[width=\textwidth]{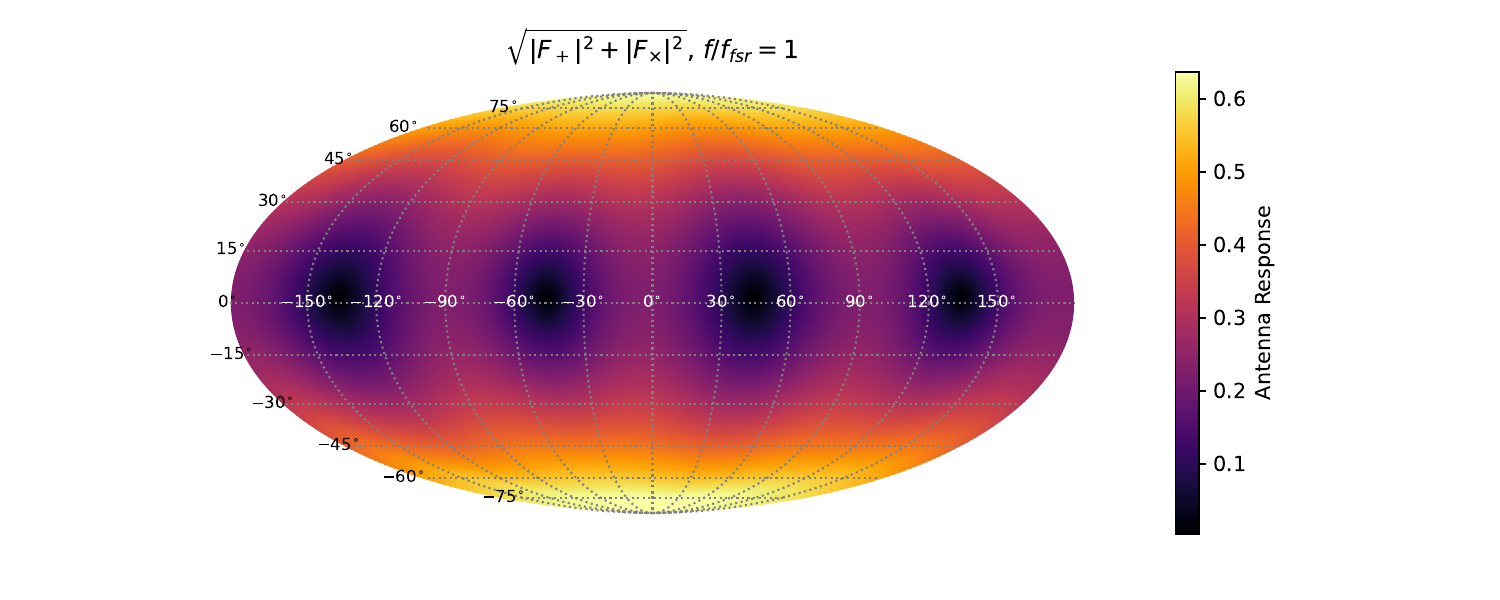}

    \end{minipage}
    \begin{minipage}[t]{0.492\linewidth}
        \centering
        \includegraphics[width=\textwidth]{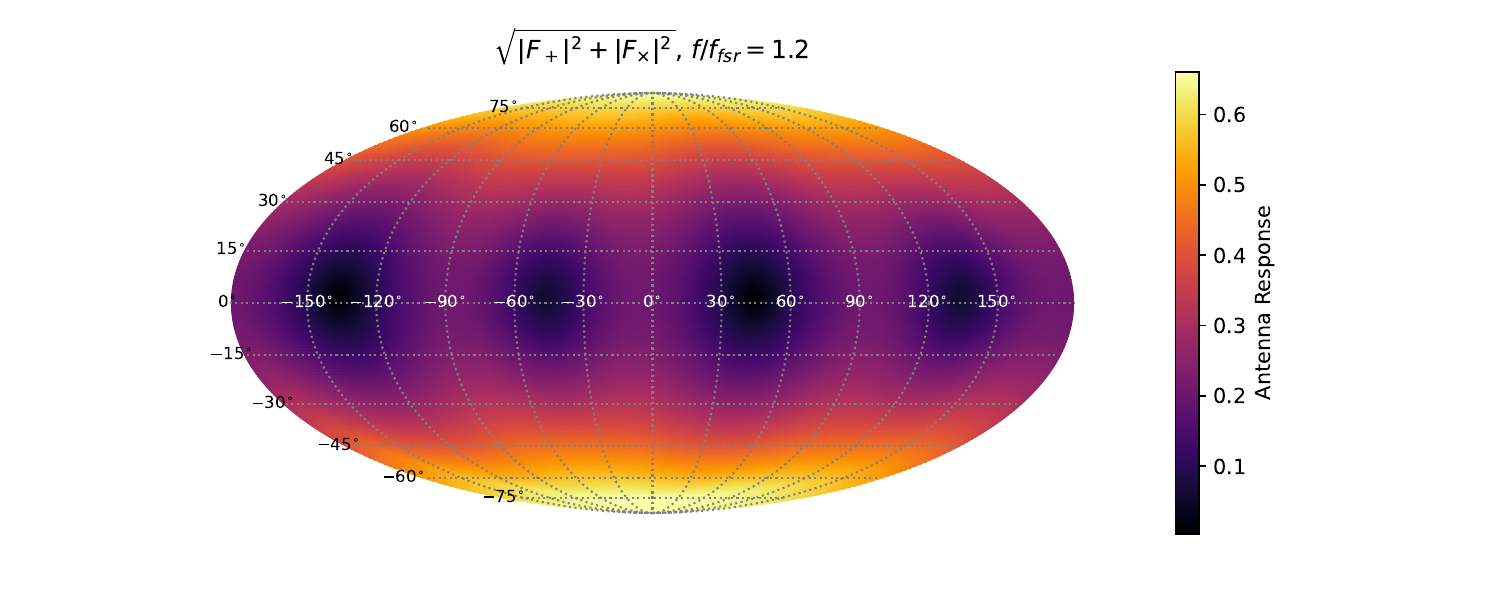}

    \end{minipage}

    \begin{minipage}[t]{0.492\linewidth}
        \centering
        \includegraphics[width=\textwidth]{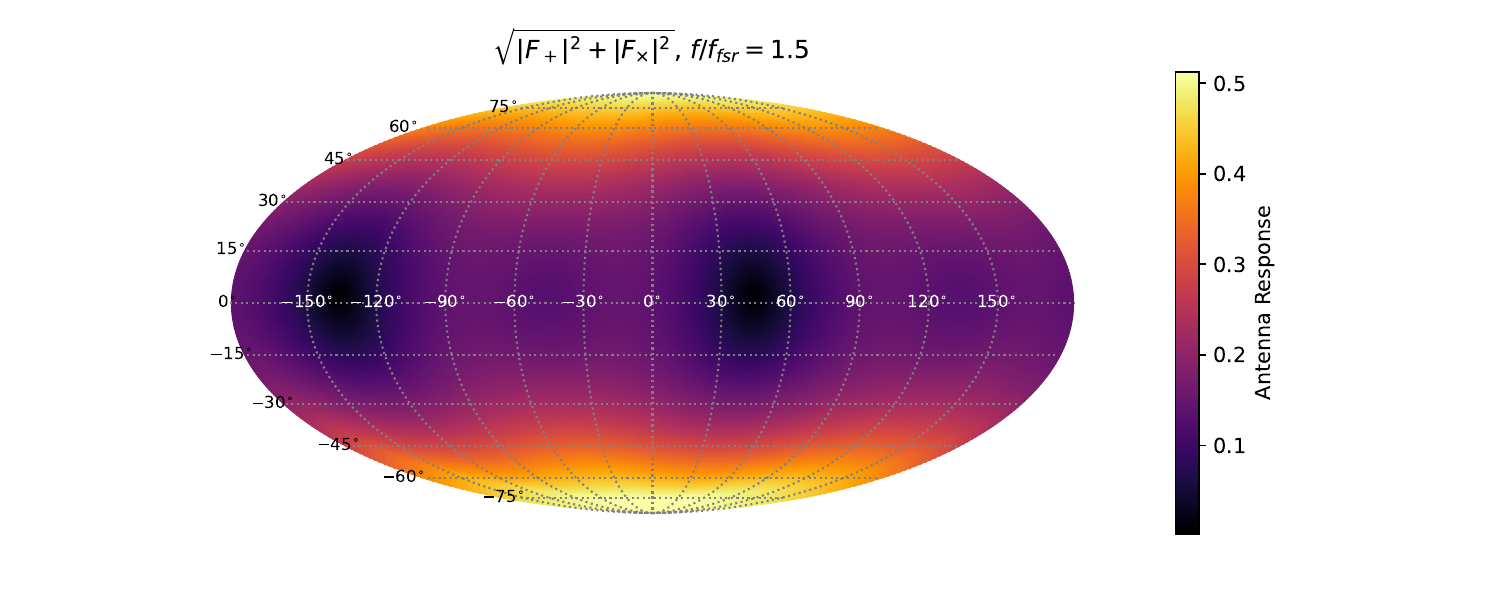}

    \end{minipage}
\caption{Spatial dependence of the magnitude of the overall antenna response factor at several specialized frequencies, which directly reflects the overall sensitivity of a gravitational wave detector aligned with the coordinate axes.}
\label{fig:spatial}
\end{figure}

The precise formulation of the antenna response in the Folded Michelson configuration may lack immediate clarity. Additionally, as the configuration works in the kilohertz band, it is impossible to linearly expand $f$ to estimate its response to gravitational waves as in traditional detectors.  Nonetheless, employing Bode plots and projection maps facilitates the visualization of the frequency-dependent variations in antenna response at specific incident angles and the spatial distribution of antenna response at particular frequencies, as shown in Fig.\,\ref{fig:freq} and Fig.\,\ref{fig:spatial} , from which we could clearly see the new configuration achieves enhanced gravitational wave detection efficiency in the vicinity of $f = f_{\text{FSR}}$ at some special angle of incidence, accompanied by a notable anisotropic spatial distribution of detection sensitivity.

\begin{figure}[htbp]
\centering

\includegraphics[width=0.48\textwidth]{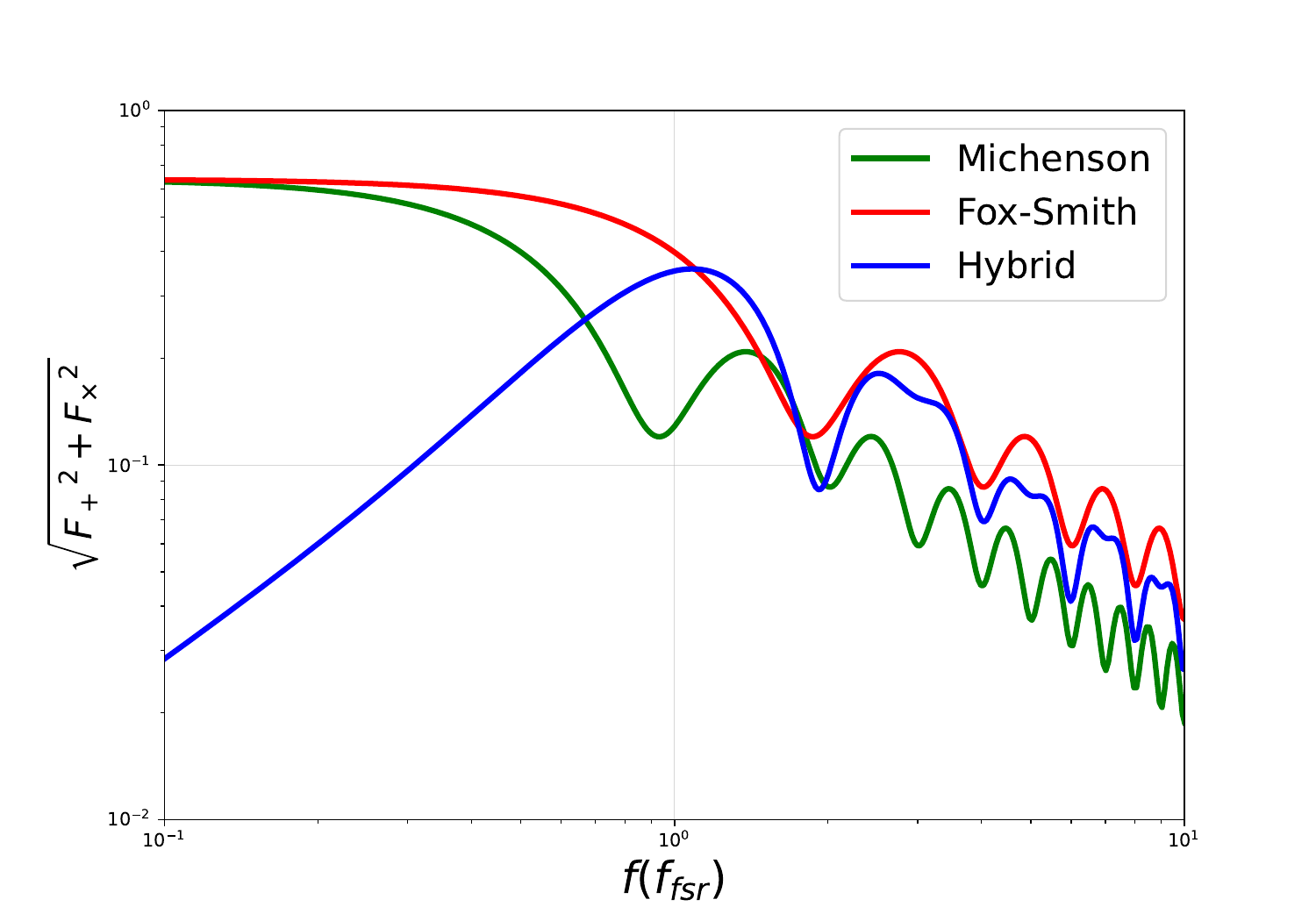}

\caption{Comparison of the sky-averaged normalised antenna response of 3 configurations.  }
\label{fig:antennares_comp}
\end{figure}

\begin{figure}[htbp]
\centering

\includegraphics[width=0.48\textwidth]{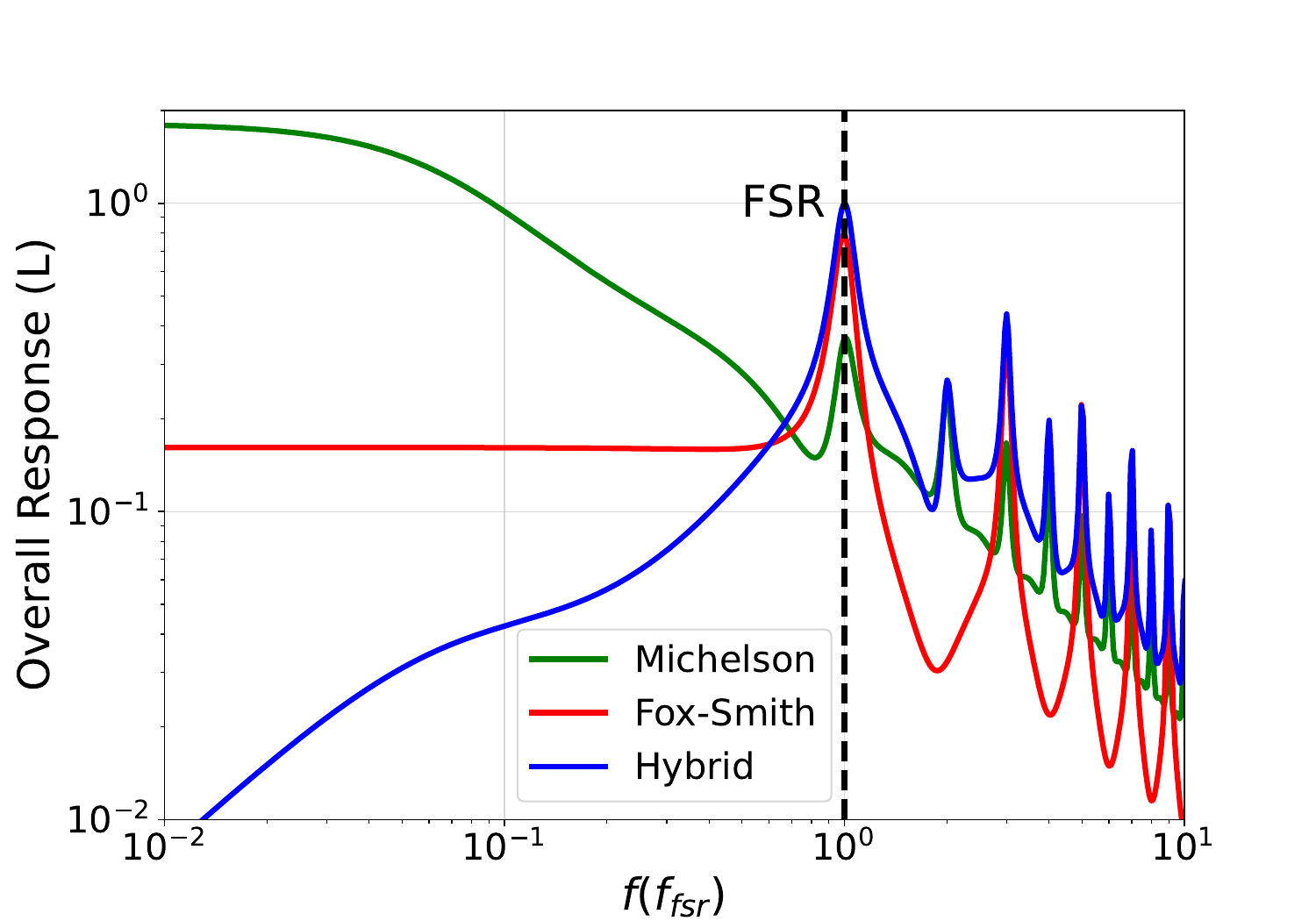}

\caption{Comparison of sky-averaged overall response to gravitational signals of three configurations, with the same circulating power and optical bandwidth of the arm cavity. }
\label{fig:ant_comp}
\end{figure}

At the same time, we also calculated the average of antenna response over the whole sky for the Hybrid design, which directly characterizes its detection capability for non-directional gravitational wave sources, such as stochastic signals\cite{MHz}. Fig.\,\ref{fig:antennares_comp} and Fig.\,\ref{fig:ant_comp} compare the average antenna response and sky-averaged sensitivity of the Hybrid design with those of the traditional Michelson and Fox-Smith configurations, which is basically a sky-averaged version of Fig.\,\ref{fig:antennares} and Fig.\,\ref{fig:ant}.

From these results, we can see that the intuitive analysis presented in Sec.\,\ref{sec:overview} was slightly exaggerated. The traditional configuration does not completely lose the ability to detect gravitational wave signals at the first FSR, and the average response efficiency of the L-shaped cavity at the first FSR is not as high as in the special case of normal incidence. Nevertheless, the L-shaped folding yields a notable improvement in sensitivity—approximately 3-5 times higher compared to the linear cavity—near the nearest optical resonance. This improvement arises because, in a sky-averaged context, the intuitive picture we introduced earlier still holds. The minimum of the averaged antenna response generally locates at frequencies corresponding to the spatial periodicity of the cavity. In the L-shaped cavity, we simply folded the cavity vertically, reducing the spatial periodicity of the arm by half. This shifts the first local minimum of the antenna response of the linear cavity, which typically appears at the first optical resonance, to the second one, directly resulting in a significant improvement in the detection capability of gravitational wave signals at the first FSR.

\newpage

\end{document}